\documentclass{ifacconf}
\usepackage{graphicx}      
\usepackage{natbib}        
\usepackage{amsmath}
\usepackage{amssymb}
\usepackage{amsfonts}
\usepackage{enumerate}
\usepackage{caption}
\usepackage{subcaption}
\begin{document}
\begin{frontmatter}

\title{Distributed payoff allocation in coalitional games via time varying paracontractions} 
\thanks[footnoteinfo]{
This work was partially supported by NWO under research project P2P-TALES (grant n. 647.003.003) and the ERC project COSMOS, (802348).Email address: \{a.a.raja, s.grammatico\}@tudelft.nl.} 
\author{Aitazaz Ali Raja,  } 
\author{  Sergio Grammatico} 
\address{Delft Center for Systems and Control, TU Delft, The Netherlands.}
\begin{abstract}                
We present a partial operator-theoretic characterization of approachability principle and based on this characterization, we interpret a particular distributed payoff allocation algorithm to be a sequence of time varying paracontractions. Further, we also propose a distributed algorithm, under the context of coalitional game, on time-varying communication network. The state in the proposed algorithm converges to a consensus within, the predefined, desired set. For convergence analysis, we rely on the operator-theoretic property of paracontraction.    
\end{abstract}


\end{frontmatter}

\section{Introduction}
Coalitional game theory provides an analytical framework and mathematical formalism, to study the behavior of selfish and rational agents, when they are willing to cooperate. Interestingly, this scenario arises in many applications, such as demand side energy management [\cite{han2018incentivizing}], in power networks for transmission cost allocation  [\cite{zolezzi2002transmission}] and cooperation between microgrids in distribution networks [\cite{Saad2011}], in various areas of communication networks by [\cite{Saad2009}], [\cite{Saad2009a}] and as conceptual foundation for coalitional control [\cite{fele2017coalitional}].\\ 
Specifically, a coalitional game with \textit{transferable utility} consists of a set of agents referred as players, who can form coalitions, and a characteristic function that determines the \textit{value} of each coalition. Note that a selfish agent will cooperate with other agents only if this coalition results in increasing its own benefit. The latter is determined by the payoff the agent receives from the value generated by a coalition. The design of criteria for determining this payoff has received acute attention by research community, such as \cite{scarf1967core}, \cite{shapley1953value}, \cite{schmeidler1969nucleolus}, \cite{maschler1971kernel}. The solutions proposed determine the stability of a coalition, i.e., whether the coalition remains intact or gets defected by its agents. One of the most widely studied solution concepts is the CORE which ensures the \textit{stability} of a game.\\ 
The problem we address in this paper is finding a payoff that belongs to CORE and hence encourages cooperation. Our practical treatment of this problem is in a multi-agent scenario, where players interact autonomously and in distributed manner to arrive at common agreement on  a payoff vector in the CORE. In this direction, \cite{lehrer2003allocation} presented an allocation process which converges to the CORE (or if this is empty, to a least-CORE). \cite{smyrnakis2019game} also consider an allocation process but under noisy observations and dynamic environment. \cite{bauso2014robust} provide conditions for an averaging process, with dynamics subject to controls and adversarial disturbances, under which the allocations converge to consensus in the desired set. \cite{Nedic2013} propose an elegant distributed bargaining algorithm which converges to a random CORE payoff allocation. The key inspiration, however, of our work is the distributed payoff allocation algorithm proposed by \cite{Bauso2015}. Their algorithm is based on the approachability principle, which is a geometric condition introduced in Blackwell's approachability Theorem [\cite{Blackwell1954}]. The approachability principle provides a way to approach a particular set and hence can be exploited to reach the CORE in the context of coalitional game theory. 

\textit{Contribution}:
In this paper, we first show that the approachability condition contains a paracontraction operator. Briefly, an operator $T: \mathbb{R}^n \to \mathbb{R}^n$ is said to be a paracontraction if, for any fixed point $y=T(y)$ and $x \in \mathbb{R}^n$, where $x \neq y$, it holds that $\| T(x) - y \| < \| x - y \|$. These operators form the subclass of, perhaps more known, \textit{quasi-non-expansive} mappings [\cite{ErnestRyu2016}].\\
Secondly, we propose a distributed payoff allocation algorithm, in context of coalitional games over time-varying communication networks. The state of proposed algorithm converges to a consensus vector that belongs to the CORE. Our approach to prove convergence of our algorithm relies on the paracontraction property of the adopted operator. 

\textit{Organization of the paper}: In Section \ref{sec: prob setup}, we provide the mathematical background for coalitional games and distributed allocation process. In Section \ref{sec: dist payoff alloc}, we discuss the approachability principle and recall the distributed payoff allocation algorithm by \cite{Bauso2015}. In Section \ref{sec: op_th}, we provide a partial operator-theoretic characterization of the approachability principle,  and we discuss algorithm in [\cite{Bauso2015}]. In Section \ref{sec: Our Algo}, we propose an algorithm for distributed allocation in coalitional games and establish its convergence using operator-theoretic properties. Further, we asses the convergence speed of proposed algorithm in Section \ref{sec: simulations}, and in Section \ref{sec: conclusion}, we conclude the paper. 

\textit{Notation}:
$\mathbb{R}$ and $\mathbb{N}$ denote the set of real and natural numbers, respectively. Given a mapping $M: \mathbb{R}^n \rightarrow \mathbb{R}^n, \mathrm{fix}(M):= \{x \in \mathbb{R}^n \mid x = M(x)\} $ denote the set of fixed points. $\text{Id}$ denotes the identity operator. For a closed set \(S \subseteq \mathbb{R}^{n},\) the mapping $\mathrm{proj}_S$: \(\mathbb{R}^{n} \rightarrow S\) denotes the projection onto \(S,\) i.e., \(\operatorname{proj}_{S}(x)=\) \(\arg \min _{y \in S}\|y-x\| .\) \(A \otimes B\) denotes the Kronecker product between the matrices \(A\) and \(B .\) $I_N$ denotes an identity matrix of dimension $N \times N$. $\mathrm{dist}(x,S)$ denotes the distance of $x$ from a closed set \(S \subseteq \mathbb{R}^{n},\) i.e., $\mathrm{dist}(x,S):= \mathrm{inf}_{y \in S} \|y-x\|$.

\section{Mathematical Background on coalitional games}\label{sec: prob setup}
A coalitional game consists of a set of agents, indexed by $\mathcal{I} = \{1, \ldots, N\}$, who cooperate to achieve selfish interests. This cooperation results in generation of utility as defined by the characteristic function $v$. Formally,
\begin{defn} (Coalitional game):
A transferable utility (TU) coalitional game is a pair  $G = (\mathcal{I},v)$, where $\mathcal{I} = \{1, \ldots, N\}$ is the index set of the agents and  $v:2^{N} \rightarrow \mathbb{R} $ is a characteristic function which assigns a real value, $v(S)$, to each coalition $S \subseteq \mathcal{I} $. $v(\mathcal{I})$ is the value of so-called grand coalition. By convention, $v(0) = \varnothing$. $\hfill \square$
\end{defn} 
The idea of coalitional game is that the value attained by a coalition $S$, i.e. ,$v(S)$ has to be distributed among the members of the coalition, thus each agent receives a certain payoff.
\begin{defn} \label{def: payoff} (Payoff vector):
Let $S \subseteq \mathcal{I}$ be a coalition of coalitional game $(\mathcal{I},v)$. A payoff vector is a vector $\boldsymbol{x} \in \mathbb{R}^{|S|}$. Where $x_i$ represents the share of agent $i \in S$ of $v(S)$.$\hfill \square$
\end{defn}

Let us state two important characteristics of a payoff vector which will further help us in explaining the solution concept of a coalitional game. First, for a game with a grand coalition $\mathcal{I}$, a payoff vector $x \in \mathbb{R}^N$ is said to be efficient if $\sum_{i \in \mathcal{I}}x_i = v(\mathcal{I})$. In words, all of the value generated by grand coalition will be distributed among the agents. Second, a payoff vector is rational if for every possible coalition $S \subseteq \mathcal{I} $ we have $\sum_{i \in S}x_i \geq v(S)$. Note that this should also hold for singleton coalitions $S = \{i\}$ i.e. $x_i \geq v({i}), \forall i\in \mathcal{I} $. It means that, payoff allocated to each agent should be at least equal to what they can get individually or by forming any coalition $S$ other than $\mathcal{I}$.\\
A payoff vector which is both efficient and rational lies in the CORE. CORE is the solution concept that relates with the stability of a grand coalition. Where, the idea of stability, in this context, is based on the disinterest of agents in defecting a grand coalition. Formally,
\begin{defn} (CORE):
The CORE $\mathcal{C}$ of a coalitional game ($\mathcal{I},v$) is the following set of payoff vectors:
\begin{equation} \label{core}
  \mathcal{C} :=  \bigg\{ x \in \mathbb{R}^N \mid \sum_{i \in \mathcal{I}} x_i = v(\mathcal{I}),\sum_{i \in S} x_i \geq v(S), \forall S \subseteq \mathcal{I}  \bigg\}.  
\end{equation}$\hfill \square$
\end{defn}
Each payoff allocation that belongs to CORE stabilizes the grand coalition. It implies that no agent or coalition $S \subset \mathcal{I}$ has an incentive to defect from the grand coalition.\\
In the sequel, we deal with the grand coalition only, therefore we use the CORE $\mathcal{C}$ as the solution concept. Note from (\ref{core}) that $\mathcal{C}$ is closed and convex. We also assume the  CORE to be non-empty through out the paper. 
Next, we discuss a possible strategy of finding the payoff vector, in a coalitional game $G = (\mathcal{I}, v)$, that belongs to CORE, $\mathcal{C}$ in (\ref{core}). Centralized methods for finding a vector $ \boldsymbol{x} \in \mathcal{C}$ do not capture realistic scenarios of interaction among autonomous selfish agents. Thus, distributed methods are employed that allow agents to autonomously reach a common agreement on a payoff allocation, $ \boldsymbol{x} \in \mathcal{C}$.\\
Generally, the distributed allocation is an iterative procedure in which, at each step, an agent $i$ proposes a utility distribution $ \boldsymbol{x}_i \in \mathbb{R}^N$ by averaging the proposals of all agents and introducing an innovation factor. This procedure aspires to finally reach at a mutually agreed payoff among participating agents. Eventually the proposed utility distributions $\{x_i\}_{i \in \mathcal{I}}$ must reach consensus.
\begin{defn}\label{st_vec} (Consensus set):
The consensus set $\mathcal{A} \subset \mathbb{R}^{N^{2}}$ is defined as:
\begin{equation}\label{eq: consensus}
\mathcal{A} := \{\boldsymbol{x} = \mathrm{col}(\boldsymbol{x}_1, \ldots, \boldsymbol{x}_N) \in \mathbb{R}^{N^{2}} \mid \boldsymbol{x}_i = \boldsymbol{x}_j, \forall i,j \in \mathcal{I}\}. 
\end{equation} 
\end{defn}$\hfill \square$\\
Therefore, in this paper, we consider the problem of computing a mutually agreed, payoff allocation vector in the CORE, i.e., $\bar{\boldsymbol{x}} \in \mathcal{A} \cap \mathcal{C} $, via a iterative distributed allocation, i.e., $\boldsymbol{x}(k) \to  \bar{ \boldsymbol{x}} \;\; \text{as} \;\;\; k \to \infty.$ 

\section{Approachability principle and Distributed payoff allocation }\label{sec: dist payoff alloc}
\subsection{Approachability principle}
We now discuss a geometric principle which can guarantee the convergence of a payoff allocation sequence to a target set, which in our coalitional game theory context, is the CORE $\mathcal{C}$, as in (\ref{core}).This principle, which we refer to as approachability principle, is the geometric concept behind celebrated approachability theorem by Blackwell presented in [\cite{Blackwell1954}].\\
\begin{figure}
\centering
\includegraphics[width=5cm]{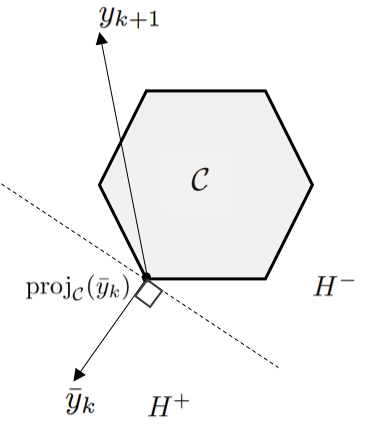}
\caption{Geometric interpretation of the approachability principle.} 
\label{fig:apr_prin}
\end{figure}
\begin{defn}(Approachability Principle)[\cite{lehrer2003allocation}, 3.2],
Let $(y_k)_{k \in \mathbb{N}}$ be a sequence of uniformly bounded vectors in $\mathbb{R}^n,$ with running average $\bar{y}_k := \frac{1}{k} \sum_{k^\prime =1}^k y_{k^\prime}$, and let $\mathcal{C}$ be a non-empty, closed and convex set. If the sequence satisfies the condition,
\begin{equation}\label{aproach}
    (\bar{y}_k - \mathrm{proj}_{\mathcal{C}}(\bar{y}_k))^\top(y_{k+1} - \mathrm{proj}_{\mathcal{C}}(\bar{y}_k)) \leq 0, \quad \forall k \in \mathbb{N},
\end{equation}
then $\displaystyle \lim_{k \to \infty} \mathrm{dist}(\bar{y}_k, \mathcal{C}) = 0.$ $\hfill \square$
\end{defn}
In Figure \ref{fig:apr_prin}, we illustrate the approachability condition in (\ref{aproach}). Let us give a geometric interpretation: the hyperplane through the point $\mathrm{proj}_{\mathcal{C}}(\bar{y}_k)$, perpendicular to the vector $(\bar{y}_k - \mathrm{proj}_{\mathcal{C}}(\bar{y}_k))$, which is the first term in (\ref{aproach}), separates the space into the half-spaces $H^+ \text{ and } H^-$. The the approachability condition requires that, the innovation $y_{k+1}$ and the average $\bar{y}_k$ should not lie in the same half-space. \\
Among others, \cite{Bauso2015} have used the approachability principle to design a distributed payoff allocation algorithm which converges to a consensus vector in the CORE in (\ref{core}). Let us recall their setup and solution algorithm in next subsection. 
\subsection{A time-varying Distributed payoff allocation process} \label{sec: DPA}
 Consider a set of agents $\mathcal{I} = \{1, \ldots, N\}$ who synchronously propose a distribution of utility at each discrete time step $k \in \mathbb{N}$. Specifically, each agent $i \in \mathcal{I}$ proposes a payoff distribution $\hat{ \boldsymbol{x}}_i(k) \in \mathbb{R}^N$, where the $j$th element denotes the share of agent $j$ proposed by agent $i$. Then, each agent $i$ computes $ \hat{\boldsymbol{x}}_i$ by averaging the proposals by his \textit{neighboring agents} and then by generating an innovation vector $ \boldsymbol{x}$ as follows:
 \begin{equation}\label{main_it}
    \hat{ \boldsymbol{x}}(k+1) = (1-\alpha_k)\boldsymbol{A}_k \hat{ \boldsymbol{x}}(k) + \alpha_k  \boldsymbol{x}(k+1), \quad \forall k \in \mathbb{N},
\end{equation}
where $(\alpha_k)_{k \in \mathbb{N}}$ is a positive sequence of step sizes, with $\alpha_k := \frac{1}{k+1}$, and $\boldsymbol{A}_k := A(k) \otimes I_N $ represents an adjacency matrix.\\
Now, Let the communication graph vary over time as $\mathcal{G}(k)=(\mathcal{I}, \mathcal{E}(k)) $. Specifically, $(j, i) \in \mathcal{E}(k)$ means that there is an active link between agents $i$ and $j$ at time $k$. In [\cite{Bauso2015}, Assumption 2], the graph sequence $(\mathcal{G}(k))_{k \in \mathbb{N}}$ is assumed to be $Q -$connected.
\begin{assum}\label{asm: Q-con }
 There exists an integer $Q \geq 1$  such that the graph $(\mathcal{I}, \cup_{l=1}^{Q} \mathcal{E}(l+k))$  is strongly connected, for all $k \geq 0 $.  $\hfill \square$
\end{assum}
The communication links in $\mathcal{G}(k)$ are weighted using an adjacency matrix $A(k) = [a_{i,j}(k)]_{N \times N}$, whose element $a_{i,j}$ represents the weight assigned by agent $i$ to the payoff distribution proposed by agent $j$, $\hat{ \boldsymbol{x}}_j(k)$. By [\cite{Bauso2015}, Assumption 1], the adjacency matrix is always doubly stochastic with positive diagonal.
\begin{assum}\label{asm: graph}
 For all $k \geq 0$, the matrix $A(k) = [a_{i,j}(k)]_{N \times N}$ satisfies following conditions:
    \begin{enumerate}[(i)]
        \item It is doubly stochastic;
        \item its diagonal elements are strictly positive, i.e., $a_{i,i}(k) > 0, \forall i \in \mathcal{I}$;
        \item $\exists$ $\gamma > 0$ such that $a_{i,j}(k) \geq \gamma$ whenever $a_{i,j}(k) > 0$. 
    \end{enumerate} $\hfill \square$
\end{assum}

Furthermore, at each time $k$, the agents generate an innovation vector $\boldsymbol{x}(k)$ in (\ref{main_it}), satisfying approachability condition, as formulated in (\ref{aproach}). Specifically, let $\boldsymbol{w}(k) := \boldsymbol{A}_k \hat{ \boldsymbol{x}}(k)$, with $\hat{ \boldsymbol{x}}(k)$ as in (\ref{main_it}), then following is postulated in [\cite{Bauso2015}, Assumption 4]: 
\begin{assum}\label{asm: bounded}
For each $k \in \mathbb{N}$, the innovation vector $ \boldsymbol{x}(k + 1)$ in (\ref{main_it}) satisfies the following inequality: 
\begin{equation}\label{cond}
\left(\boldsymbol{w}(k)-\mathrm{proj}_{\mathcal{C}}(\boldsymbol{w}(k))\right)^\top \left( \boldsymbol{x}(k+1)-\mathrm{proj}_{\mathcal{C}}(\boldsymbol{w}(k))\right) \leq 0,
\end{equation} 
where $\mathcal{C}$ is the CORE set as in (\ref{core}). $\hfill \square$
\end{assum}
Moreover, to fulfil the conditions of the approachability principle, the innovation vector is uniformly bounded, [\cite{Bauso2015}, Assumption 4].
\begin{assum}\label{asm: uniformly bounded}
Let $\boldsymbol{x}(k+1)$ be innovation vector in (\ref{main_it}). There exist $L > 0$, such that $ \|  \boldsymbol{x}_i(k + 1)\| \leq L, \forall k \geq 0$.  
\end{assum}
 The main result regarding the iteration in (\ref{main_it}) by \cite{Bauso2015} is that, if Assumptions \ref{asm: Q-con }$-$\ref{asm: uniformly bounded} hold then the average allocation vector $\hat{ \boldsymbol{x}}(k)$ will converge to the set $\mathcal{A} \cap \mathcal{C}$. In the context of coalitional game theory, this implies that through the distributed allocation process in (\ref{main_it}), the agents will reach a common agreement on the payoff distribution, which lies in the CORE.
\section{operator theoretic characterization}\label{sec: op_th}
\subsection{Approachability principle as a paracontraction}
In this subsection, we aim at providing an operator-theoretic characterization of the approachability condition in (\ref{cond}), and present an interesting operator contained by approachability condition which holds a \textit{paracontraction} property. To show that, we first define the notion of paracontraction.
\begin{defn}\label{def: para} (Paracontraction):
A continuous mapping $M : \mathbb{R}^n \rightarrow \mathbb{R}^n$ is a paracontraction, with respect to a norm $\|\cdot\|$ on $\mathbb{R}^n$, if
\begin{equation*}
    \|M(x) - y\| < \|x - y\|,
\end{equation*}
for all $x,y \in \mathbb{R}^n \text{ such that } x \notin \mathrm{fix}(M), y \in \mathrm{fix}(M)$. $\hfill \square$
\end{defn}
The approachability condition in (\ref{cond}), given $\boldsymbol{w}(k) = \boldsymbol{A}_k\hat{\boldsymbol{x}}(k)$ provides us the criterion for generating an innovation vector $\boldsymbol{x}(k+1)$ to be used in the iterative process in (\ref{main_it}). 
In the next statement, we will present an alternative formulation for the approachability condition which, interestingly, is the sum of a paracontracting operator and arbitrary vectors with specific geometric meaning.

\begin{lem} \label{lem: approachability condition}
Let $\beta \in[0,1)$, $Q_C := 2\mathrm{ proj}_C - \text{Id}$ be the over-projection operator, $\mathbf{v}^{\bot}(k) = \mathbf{v}^{\bot}(\boldsymbol{w}_i(k))$ be an arbitrary vector that belongs to the hyperplane orthogonal to the vector $  \boldsymbol{u} :=(\boldsymbol{w}_i(k)-\mathrm{proj}_\mathcal{C}(\boldsymbol{w}_i(k))) $ in (\ref{cond}) and $\mathbf{v}^-(k) = \mathbf{v}^-(\boldsymbol{w}_i(k))$ be a vector orthogonal to $\mathbf{v}^{\bot}(k)$ in the direction opposite to vector $\boldsymbol{u}$, (Figure \ref{fig: apr_applied_exp}). Then, the following equation corresponds exactly to the approachability condition in (\ref{cond}):
\begin{equation}\label{conv_para}
\begin{array}{ll}
\boldsymbol{x}_i(k+1) = (1 - \beta) \mathrm{proj}_\mathcal{C}(\boldsymbol{w}_i(k)) + &\beta Q_{\mathcal{C}}(\boldsymbol{w}_i(k)) \\
&+ \mathbf{v}^{\bot}(k)+ \mathbf{v}^-(k). \quad \hfill \square 
\end{array}
\end{equation}
\begin{figure}
\begin{subfigure}{.5\textwidth}
  \centering
  \includegraphics[width=.5\linewidth]{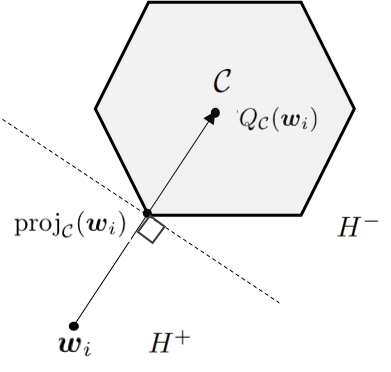}  
  \caption{}
  \label{fig:sub-first}
\end{subfigure}
\begin{subfigure}{.5\textwidth}
  \centering
  \includegraphics[width=.5\linewidth]{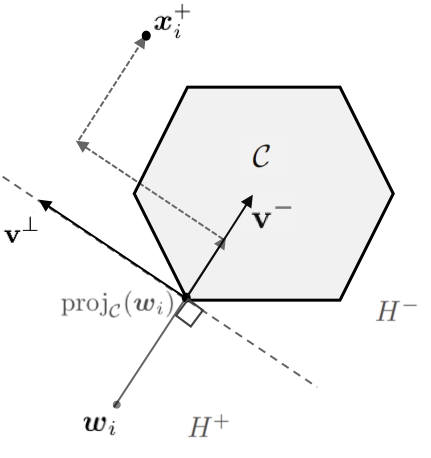}  
  \caption{}
  \label{fig:sub-second}
\end{subfigure}
\caption{Illustration of the approachability condition as in Equation (\ref{conv_para}): projection and over-projection (a);  innovation $\boldsymbol{x}_i^+$  (b).}
\label{fig: apr_applied_exp}
\end{figure}
In Figure \ref{fig: apr_applied_exp}, we geometrically illustrate Equation (\ref{conv_para}) for some $\beta \in (1/2, 1)$.
\end{lem}
\begin{pf}
To show that (\ref{conv_para}) corresponds to the approachability condition, let us plug (\ref{conv_para}) into (\ref{cond}). In the remainder of the proof, we drop the dependence on $k$ for ease of notation.

$
\begin{array}{lllll}
 \underbrace{(\boldsymbol{w}_i - \; \mathrm{proj}(\boldsymbol{w}_i))^\top}_{\boldsymbol{u}} (\underbrace{\boldsymbol{x}^+}_{(\ref{conv_para})} - \mathrm{proj}(\boldsymbol{w}_i)) &\leq 0\\
\Leftrightarrow(\boldsymbol{u})^{\top}((1 - \beta) \mathrm{proj}_\mathcal{C}(\boldsymbol{w}_i) + \beta\underbrace{Q_{\mathcal{C}}(\boldsymbol{w}_i)}_{2\mathrm{proj}_\mathcal{C}(\boldsymbol{w}_i) - \boldsymbol{w}_i} \\
\qquad \qquad \qquad \qquad \qquad + \mathbf{v}^{\bot} + \mathbf{v}^- - \mathrm{proj}(\boldsymbol{w}_i))  &\leq 0\\
\Leftrightarrow(\boldsymbol{u})^{\top}(\beta(-\boldsymbol{u}) + \mathbf{v}^{\bot} + \mathbf{v}^-) &\leq 0 \\
\Leftrightarrow -\beta (\boldsymbol{u})^{\top}(\boldsymbol{u}) + \underbrace{(\boldsymbol{u})^{\top}\mathbf{v}^{\bot}}_0
+ (\boldsymbol{u})^{\top}\mathbf{v}^- &\leq 0 \\
\Leftrightarrow -\beta \|\boldsymbol{u}\|^2 - |\boldsymbol{u}|\; |\mathbf{v}^-| &\leq 0,\\
\end{array}$\\
Since all the steps are equivalent and the vectors $ \mathbf{v}^{\bot}$ and $\mathbf{v}^-$ can be chosen arbitrarily for each given $\boldsymbol{w}_i(k)$, and since any point in $H^-$ can be written in the form in (\ref{conv_para}), we conclude that (\ref{conv_para}) is equivalent to the approachability condition in (\ref{cond}). $\hfill \blacksquare$
\end{pf}
Let us now consider the particular case of (\ref{conv_para}) with $ \mathbf{v}^{\bot} = \mathbf{v}^- = 0$, and define the dependence of $\boldsymbol{x}(k+1)$ from $\boldsymbol{w}(k)$ via an operator $\mathcal{T}$: 
\begin{equation}\label{eq: apr special case}
\begin{array}{lll}
    \forall i \in \mathcal{I}: \boldsymbol{x}_i^+ &= (1 - \beta) \mathrm{proj}_\mathcal{C}(\boldsymbol{w}_i) + \beta Q_{\mathcal{C}}(\boldsymbol{w}_i)\\
        &= \underbrace{((1 - \beta) \mathrm{proj}_\mathcal{C}+ \beta Q_{\mathcal{C}})}_{\mathcal{T}_i}(\boldsymbol{w}_i)\\
        &= \mathcal{T}_i(\boldsymbol{w}_i).
\end{array}
\end{equation}
The operator $\mathcal{T}_i:= (1 - \beta) \mathrm{proj}_\mathcal{C}(\cdot) + \beta Q_{\mathcal{C}}(\cdot)$ in  (\ref{eq: apr special case}) is a mapping from $\boldsymbol{w}(k)$ to $\boldsymbol{x}(k+1)$ which, by Lemma \ref{lem: approachability condition}, satisfies the approachability condition in (\ref{cond}). Using this operator $\mathcal{T}_i$, we can give the following representation to the process of generation of an innovation vector $\boldsymbol{x}(k+1)$ in (\ref{main_it}), which is equivalent to the particular case in (\ref{eq: apr special case}) of the approachability condition.
\begin{equation}\label{eq: opr T}
  \boldsymbol{x}(k+1) = \mathcal{T}(\boldsymbol{w}(k)) = \begin{bmatrix}
\mathcal{T}_1(w_1(k))\\
\vdots\\
\mathcal{T}_N(w_N(k))
\end{bmatrix}. 
\end{equation}
Next, we present an operator-theoretic property of the operator $\mathcal{T}$ in the following statement.
\begin{thm}\label{apr_para}
The operator $\mathcal{T}: \mathbb{R}^n \to \mathbb{R}^n$ defined in (\ref{eq: apr special case})$-$(\ref{eq: opr T}) is a paracontraction. $\hfill \square$
\end{thm}
Before presenting the proof of Theorem \ref{apr_para}, we provide two technical statements, which we exploit later in the proof.
\begin{lem}\label{lem: proj, over proj} (Projection and Over-projection operators):
Let $C \subset \mathbb{R}^n$ be a non-empty, closed and convex set. Then, with respect to the Euclidean norm $\|\cdot\|_2$:
\begin{enumerate}[(i)]
    \item the projection operator $\mathrm{proj}_C$ is a paracontraction;
    \item the over projection operator, $Q_C := 2\mathrm{ proj}_C - \text{Id}$, is non-expansive. $\hfill \square$
\end{enumerate} 

\end{lem}
\begin{pf} (i): If $C$ is closed and convex then $\mathrm{proj}_C$ is a paracontraction, [\cite{Elsner1992}, Example 2]. \\
(ii): By [\cite{ErnestRyu2016}, Subsection 3.1].
$\hfill \blacksquare$
\end{pf}
\begin{lem}\label{convex_comb}
Let $M$ be a paracontraction, $B$ be a non-expansive operator, with $\mathrm{fix}(M)  \cap  \mathrm{fix}(B) \neq \varnothing$ and $\alpha \in (0,1)$. Then, $C:=(1-\alpha)M + \alpha B$ is a paracontraction. $\hfill \square$
\end{lem}
\begin{pf}
Let $y \in \mathrm{fix}(M) \cap \mathrm{fix}(B)$ and $x \neq y$. Then:
\begin{flalign*}
\|C(x) -  C(y)\| &= \|((1-\alpha)M + \alpha B)x \\
 &\qquad \qquad \qquad \qquad - ((1-\alpha)M + \alpha B)y\| \\
&= \|(1-\alpha)(Mx - My) + \alpha(Bx - By)\|\\
&\leq (1-\alpha) \|Mx - y\| + \alpha\|(Bx - y)\|\\
& < (1-\alpha) \|x - y\| + \alpha\|(x - y)\|\\
& = \|x - y\|,
\end{flalign*}
where  we have used the triangular inequality and then the definition of paracontraction for $M$. Therefore, with $\|C(x) -  C(y)\| < \|x - y\|$, we obtain the definition of paracontraction.    $\hfill \blacksquare$ 
\end{pf}
\begin{rem}\label{rem: both_para}
Lemma \ref{convex_comb} also holds if both operators are paracontractions (with the same proof). $\hfill \square$
\end{rem}
Given these results, we are now ready to present the proof of Theorem \ref{apr_para}.

\begin{pf}(Theorem \ref{apr_para}):
At each time $k$ an agent $i$ generates an innovation vector $ \boldsymbol{x}_i(k+1)$ in (\ref{main_it}), satisfying the restricted approachability condition in (\ref{eq: apr special case}).
By Lemma \ref{lem: proj, over proj}, the operator $\mathcal{T}$ in (\ref{eq: opr T}) is a convex combination of a paracontraction, $\mathrm{proj}_\mathcal{C}(\cdot)$ and a non-expansive operator,  $Q_\mathcal{C}(\cdot)$. Thus, by Lemma \ref{convex_comb}, it is a paracontraction. $\hfill \blacksquare$
\end{pf}
\subsection{Distributed allocation process as a sequence of time varying paracontractions}
The result in Theorem \ref{apr_para} further allows us to characterize an operator-theoretic property of the iteration in (\ref{main_it}). We show that, under a particular case of approachability condition in (\ref{eq: apr special case}), the iteration generates a sequence of time varying paracontractions. To prove this, we recall two useful results related to paracontractions.
\begin{prop}\label{p1} (Composition of paracontracting operators):
Suppose $M_1, M_2 : \mathbb{R}^n \rightarrow \mathbb{R}^n$ are paracontractions with respect to same norm $\|\cdot\|$ and $\mathrm{fix}(M_1) \cap \mathrm{fix}(M_2) \neq \emptyset$. Then the composition $M_1 \circ M_2$ is a paracontraction with respect to the norm $\|\cdot \|$ and $\mathrm{fix}(M_1 \circ M_2) = \mathrm{fix}(M_1) \cap \mathrm{fix}(M_2)$, [\cite{Fullmer2018}, Prop. 1]. $\hfill \square$
\end{prop}

\begin{prop}\label{p3} (Doubly stochastic matrix): 
  Let $A$ be a doubly stochastic matrix with strictly positive diagonal elements. Then, the linear operator defined by the matrix $A \otimes I_{n} $ is a paracontraction with respect to the mixed vector norm $\|\cdot\|_{2,2}$, [\cite{Fullmer2018}, Prop. 5]. $\hfill \square$
\end{prop}
Using the operator $\mathcal{T}$ in (\ref{eq: opr T}) and $\boldsymbol{w}(k) = \boldsymbol{A}_k \hat{ \boldsymbol{x}}(k)$ as in (\ref{cond}),  we can rewrite (\ref{main_it}) as: 
\begin{equation}\label{main_it_op}
   \boldsymbol{w}(k+1) = (1-\alpha_k)\boldsymbol{A}_k \boldsymbol{w}(k) + \alpha_k \boldsymbol{A}_k \mathcal{T}(\boldsymbol{w}(k)), \quad \forall k \in \mathbb{N}.
\end{equation}
Note that, the step-size sequence $(\alpha_k)_{k \in \mathbb{N}}$ in (\ref{main_it}) is specified to be $\alpha_k = \frac{1}{k+1}$ by \cite{Bauso2015}. Here, we can generalize it subject to the following assumption.  
\begin{assum}\label{asm: step size}
 Let $(\alpha_k)_{k > 0}$ be a sequence such that $\alpha_k \in (0,1), \forall k \geq 0$, \(\sum_{k=0}^{\infty} \alpha_k = \infty\), and \(\sum_{k=0}^{\infty} \alpha_k^2 < \infty\). $\hfill \square$
\end{assum}

Let us also define an operator  $\mathcal{S}_k := (1-\alpha_k)\boldsymbol{A}_k (\cdot)  + \alpha_k \boldsymbol{A}_k \mathcal{T}(\cdot)$, which in turn allows us to represent the iteration in ($\ref{main_it_op}$) more concisely as:
\begin{equation}\label{eq: main_it_simp}
 \boldsymbol{w}(k+1) = \mathcal{S}_k(\boldsymbol{w}(k)).
\end{equation}
With the latter formulation, we can now conveniently characterize the paracontraction property of the operator $\mathcal{S}_k$, according to the corollary below. 
\begin{cor}\label{cor: tv_para}
Let the operator $\mathcal{T}: \mathbb{R}^n \to \mathbb{R}^n$ be as in (\ref{eq: opr T}). Then, for each $k \in \mathbb{N}$, the operator $\mathcal{S}_k$ in (\ref{eq: main_it_simp}) is a paracontraction. $\hfill \square$
\end{cor}
\begin{pf}
By Theorem \ref{apr_para}, the operator $\mathcal{T}$ is a paracontraction. Furthermore, by Proposition \ref{p1} and \ref{p3}, the composition $\boldsymbol{A}_k \circ \mathcal{T}(\cdot)$ is also a paracontraction. This fact and Proposition \ref{p3} imply that for each $k \in \mathbb{N}$ the operator $\mathcal{S}_k := (1-\alpha_k)\boldsymbol{A}_k (\cdot)  + \alpha_k \boldsymbol{A}_k \mathcal{T}(\cdot)$  is a convex combination of paracontractions and hence, by Remark \ref{rem: both_para} on Lemma \ref{convex_comb}, is a paracontraction. $\hfill \blacksquare$
\end{pf} 
\begin{rem}\label{rem: alpha fixed}
Corollary \ref{cor: tv_para} also holds if, for all $k \in \mathbb{N}, \alpha_k = \alpha \in (0,1)$ in (\ref{main_it_op}). $\hfill \square$
\end{rem}
The results in Theorem \ref{apr_para} and Corollary \ref{cor: tv_para} provide an interesting operator-theoretic insight into the structure of algorithm presented by \cite{Bauso2015}. We use this insight to design our own distributed payoff allocation algorithm, which we 
present in the next section along with its convergence proof.
\section{Distributed allocation via paracontraction operators over time-varying networks}\label{sec: Our Algo}
In this section, we present our distributed allocation algorithm and exploit the results derived in Section \ref{sec: op_th} to prove its convergence. The algorithm we propose is similar, in structure, to iteration presented in (\ref{main_it_op}), so the same definitions hold except for the step size $\alpha$, which is considered to be fixed here. In fact, the paracontraction property of the employed operator in proposed algorithm, allows us to prove the convergence, even with the fixed $\alpha$.  Further, we will show in Section \ref{sec: simulations} via numerical simulations that the algorithm actually performs faster with an appropriate choice of fixed step size $\alpha$.\\
Let the elements of the iteration, i.e., the set of agents $\mathcal{I}$, the operator $\mathcal{T}$,  the vector $\boldsymbol{w}(k)$ and the matrix $\boldsymbol{A}_k = A(k) \otimes I_{N}$ be as in (\ref{main_it_op}), defined in Subsection \ref{sec: DPA}. Then, the distributed allocation procedure on time varying networks, takes the form:
\begin{equation}\label{main_it_algo}
    \boldsymbol{w}(k+1) = (1-\alpha)\boldsymbol{A}_k  \boldsymbol{w}(k) + \alpha \boldsymbol{A}_k \mathcal{T}( \boldsymbol{w}(k)), \quad \forall k \in \mathbb{N}.
\end{equation}
Note that, in our proposed iteration in (\ref{main_it_algo}), there are two differences compared to (\ref{main_it_op}). First, the step size $\alpha$ is fixed and secondly the elements of communication matrix $A(k)$ can take values from finite set. The latter implies that there are finite number of adjacency matrices available, for the communication among agents. Formally,
\begin{assum}\label{asm: fixed graph}
 Each element of communication matrix $A_k$, i.e., $a_{i,j}(k), \forall (i,j) \in \mathcal{I} $, can take the values in a finite set. $\hfill \square$
\end{assum}
We can also redefine the operator $\mathcal{S}$ in (\ref{eq: main_it_simp}) with fixed $\alpha$ as $\mathcal{S}_k := (1-\alpha)\boldsymbol{A}_k (\cdot)  + \alpha \boldsymbol{A}_k \mathcal{T}(\cdot)$ to write (\ref{main_it_algo}) in compact form as:
\begin{equation}\label{main_com}
     \boldsymbol{w}(k+1) = \mathcal{S}_k( \boldsymbol{w}(k)), \qquad \forall k \in \mathbb{N}.
\end{equation}
Note that, because of fixed step size $\alpha$ in (\ref{main_it_algo}) and Assumption \ref{asm: fixed graph}, the operator sequence $(\mathcal{S}_k)_{k \in \mathbb{N}}$ will belong to a finite family of paracontractions. This will allow us to exploit the following well-known theorem, proved by \cite{Elsner1992}, later for our convergence result.  
\begin{lem}\label{t1}(\cite{Elsner1992})
Let $\mathcal{M}$ be a finite family of paracontractions such that $\bigcap_{M \in \mathcal{M}} \mathrm{fix}(M) \neq \varnothing $, and consider the iteration
\begin{equation*}
   x(k+1) = M_k(x(k)),
\end{equation*}
where, for each $k \in \mathbb{N}, M_k \in \mathcal{M}$. Then, the state $x(k)$ converges to a common fixed point of the paracontractions that occur infinitely often in the sequence. $\hfill \square$
\end{lem}

We now have the necessary tools and algorithmic setup to show, in the following theorem, that iteration in (\ref{main_it_algo})$-$(\ref{main_com}) converges to a consensus vector, see $\mathcal{A}$ in (\ref{eq: consensus}) which belongs to CORE, $\mathcal{C}$ in (\ref{core}).
\begin{thm}
Let $\alpha \in (0,1]$ and the operator $\mathcal{T}: \mathbb{R}^n \to \mathbb{R}^n$ be a paracontraction with $\mathrm{fix}(\mathcal{T}) = \mathcal{C}$ in (\ref{core}). Let Assumptions $1-3$ and $6$ hold. Then, the iteration in (\ref{main_com}) is such that:
\begin{enumerate}[(i)]
    \item $(\mathcal{S}_k)_{k \in \mathbb{N}}$ is a sequence of time-varying paracontractions;
    \item $\mathrm{fix}(\mathcal{S}_k) = \mathcal{C} \cap \mathcal{A}, \forall k \in \mathbb{N}$;
    \item $\displaystyle \lim_{k \to \infty} \boldsymbol{w}_k = \boldsymbol{w}^* $ for some $ \boldsymbol{w}^* \in \mathcal{C} \cap \mathcal{A}$,
\end{enumerate}
where $\mathcal{C}$ is the CORE set (\ref{core}) and $\mathcal{A}$ is the consensus set (\ref{eq: consensus}). $\hfill \square$
\end{thm} 
\begin{pf}
 (i): It follows directly from Remark \ref{rem: alpha fixed} on Corollary \ref{cor: tv_para}.\\
(ii): To characterise the fixed points of $\mathcal{S}_k$ in (\ref{main_com}), let $\mathbf{y} \in \mathrm{fix}(\mathcal{S}_k)$ i.e. $\mathbf{y} = (1-\alpha)\boldsymbol{A}_k \mathbf{y} + \alpha \boldsymbol{A}_k \mathcal{T}(\mathbf{\mathbf{y}})$. And let $\bar{ \boldsymbol{w}} \in \mathrm{fix}(\mathcal{T}) \cap \mathcal{A}$. Here, we want to show that $\mathbf{y} = \bar{ \boldsymbol{w}}$. \\
It follows by Perron-Frobenius theorem  that $\mathrm{fix}(\boldsymbol{A}_k) = \mathcal{A}$, regardless of the temporal variation in $\boldsymbol{A}_k$. So, $\bar{ \boldsymbol{w}} \in \mathcal{A}\Rightarrow \bar{ \boldsymbol{w}} = \boldsymbol{A}_k \circ \mathcal{T}(\bar{ \boldsymbol{w}})$. Consequently, $\mathbf{y} = (1-\alpha)\boldsymbol{A}_k \bar{ \boldsymbol{w}} + \alpha \boldsymbol{A}_k \mathcal{T}(\bar{ \boldsymbol{w}}) \Rightarrow \mathbf{y} = \bar{ \boldsymbol{w}}$. And, as $\mathrm{fix}(\mathcal{T}) = \mathcal{C}$, hence $\mathrm{fix}(\mathcal{S}_k) = \mathcal{C} \cap \mathcal{A}$, which concludes the proof of this assertion. \\
(iii): It follows from assertion (i), (ii), Assumption \ref{asm: fixed graph} and direct application of Lemma \ref{t1}. $\hfill \blacksquare$
\end{pf} 
This result shows a remarkable ability of operator-theoretic tools to describe algorithms in general form. For instance, our algorithm in (\ref{eq: main_it_simp}) allows a mechanism designer to choose an operator $\mathcal{T}$ in (\ref{main_it_algo}) of his choice to possibly steer the consensus towards a particular point in set $\mathcal{C}$ in (\ref{core}). This operator is primarily required to fulfill two necessary requirements: it should be a paracontraction and $\mathrm{fix}(\mathcal{T}) = \mathcal{C} $.    
\begin{figure}
\begin{subfigure}{.5\textwidth}
  \centering
  \includegraphics[width=1\linewidth]{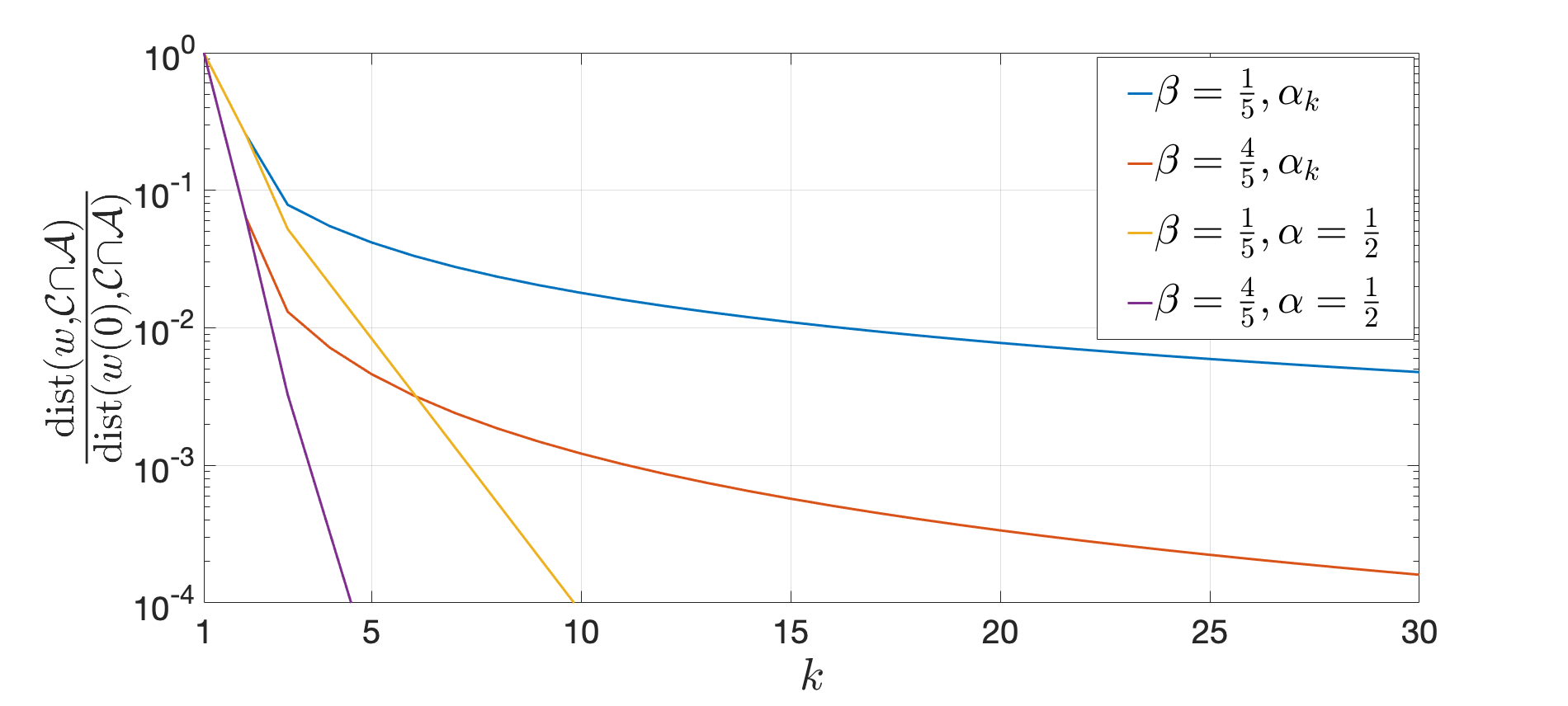}  
  \caption{}
  \label{fig:sub-one}
\end{subfigure}
\begin{subfigure}{.5\textwidth}
  \centering
  \includegraphics[width=1\linewidth]{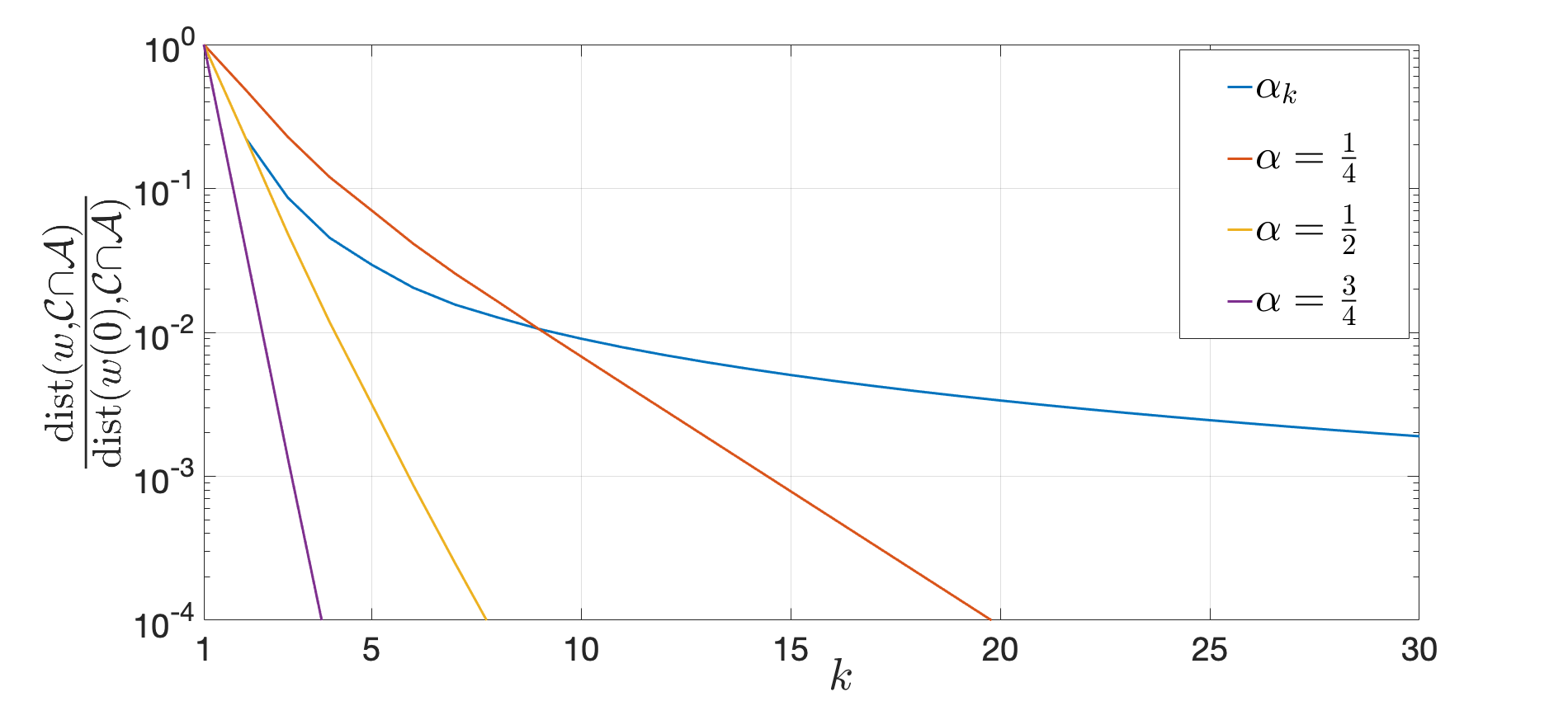}  
  \caption{}
  \label{fig:sub-two}
\end{subfigure}
\caption{(a): The trajectories of $\mathrm{dist}(\boldsymbol{w}(k), \mathcal{C}\cap \mathcal{A})/\mathrm{dist}(\boldsymbol{w}(0), \mathcal{C}\cap \mathcal{A})$ with $\alpha = 1/(k+1) \text{ for } \beta = 1/5, 4/5 \text{ and } \alpha = 1/2 \text{ for } \beta = 1/5, 4/5.$ (b): The trajectories of $\mathrm{dist}(\boldsymbol{w}(k), \mathcal{C}\cap \mathcal{A})/\mathrm{dist}(\boldsymbol{w}(0), \mathcal{C}\cap \mathcal{A})$ with $\alpha = 1/4, \alpha = 1/2, \alpha = 3/4 \text{ and } \alpha = 1/(k+1)$.}
\label{fig: results}
\end{figure}
\section{Numerical Simulations}\label{sec: simulations}
In our numerical simulations, we consider a coalitional game $(\mathcal{I},v)$ played among $N = 4$ agents with a set of agents as $\mathcal{I} = \{1,2,3,4\}$. Coalitions, including the singleton, are assigned with a value specified by characteristic function $v$. We set, $v(\{1\}) = 4, v(\{2\}) = 3, v(\{3\}) = 0, v(\{4\}) = 3, v(\{1,2\}) = 5, v(\{3,4\}) = 3, v(\{1,2,3\}) = 7, v(\mathcal{I})=10 $. Now, a payoff vector, as in Definition \ref{def: payoff}, that belongs to CORE, $\mathcal{C}$ in (\ref{core}) must allocate each agent at-least its individual value, sum of their allocations should be $v_N = 10$ and be group rational. Consistent with these requirements, the CORE of this game is the following set:
\begin{equation*}
    \begin{array}{lll}
         \mathcal{C} = \bigg \{ &x \in \mathbb{R}^4 \mid x_1 + x_2 + x_3 + x_4 = 10, \\
         & x_1 + x_2 + x_3 \geq 7, x_1 + x_2 \geq 5,\\
         &x_3 + x_4 \geq 3, x_1 \geq 4, x_2 \geq 3, x_3 \geq 0, x_4 \geq 3 \bigg \}.
    \end{array}
\end{equation*}
The agents communicate over time-varying graphs associated with the adjacency matrices $A(k)$. Here, we set the adjacency matrices to be:\\ 
$$A(2k + 1)=\left[\begin{array}{cccc}\tfrac{1}{2}  & \tfrac{1}{2} & {0}& {0}\\ \tfrac{1}{2}  & \tfrac{1}{2}  & {0}&{0} \\ {0} & {0} & \tfrac{1}{2}  & \tfrac{1}{2} \\{0} & {0} & \tfrac{1}{2} &\tfrac{1}{2} 
\end{array}\right], \;\;
 A(2k + 2)=\left[\begin{array}{cccc}\tfrac{1}{2} & {0} & \tfrac{1}{2}& {0}\\ {0} & \tfrac{1}{2} & {0}&\tfrac{1}{2} \\ \tfrac{1}{2} & {0} & \tfrac{1}{2}& {0}\\{0} & \tfrac{1}{2} & {0}&\tfrac{1}{2}
\end{array}\right],$$\\
for all $k \in \mathbb{N}$. Note that this graph sequence  satisfies Assumption \ref{asm: Q-con } with $Q =2$,  and the elements of the adjacency matrices satisfy Assumption \ref{asm: graph} with $\gamma = 1/2.$  \\
For the initial assignments, we assume that each agent allocates entire value of coalition, i.e., $v(\mathcal{I}) = 10$ to itself. For example, the initial proposal by agent 1 will be $\boldsymbol{w}_1(1) = [\:10 \; 0 \; 0 \; 0\:]^\top$. Finally, we apply the iteration in (\ref{main_it_algo}) with the operator $\mathcal{T}= (1 - \beta) \mathrm{proj}_\mathcal{C}(\cdot) + \beta Q_{\mathcal{C}}(\cdot)$, and as expected, the local allocations converge to $\mathcal{C} \cap \mathcal{A}$. \\
In Figure \ref{fig: results}(a), we compare the trajectories of normalized distances $\mathrm{dist}(\boldsymbol{w}(k), \mathcal{C}\cap \mathcal{A})/\mathrm{dist}(\boldsymbol{w}(0), \mathcal{C}\cap \mathcal{A})$, by varying $\beta$ for a specified $\alpha_k$. We can observe that a higher value of $\beta$ corresponds to a faster convergence. In Figure \ref{fig: results}(b), we use the same metric and observe the convergence speed while varying $\alpha$. As expected, the convergence of iteration with fixed step size $\alpha$, is faster compared to a decreasing sequence $(\alpha_k)_{k \in \mathbb{N}}$ as in [\cite{Bauso2015}].
\section{Conclusion}\label{sec: conclusion}
We presented a partial operator-theoretic characterization of the approachability principle and showed that it contains a paracontraction operator. Based on this result, we have proposed a distributed payoff allocation algorithm, with fixed step sizes, and proved its convergence via operator-theoretic arguments. Such analysis of algorithms, based on operator theory, allow more general description of their structure and hence open further improvement possibilities. \\
As future work, we aim to completely characterize the approachability principle in operator-theoretic terms. It would also be valuable to relax the assumption on the communication graph from double stochasticity to row stochasticity.  

\bibliography{ifacconf}             

\begin{thebibliography}{19}
\providecommand{\natexlab}[1]{#1}
\providecommand{\url}[1]{\texttt{#1}}
\providecommand{\urlprefix}{URL }
\expandafter\ifx\csname urlstyle\endcsname\relax
  \providecommand{\doi}[1]{doi:\discretionary{}{}{}#1}\else
  \providecommand{\doi}{doi:\discretionary{}{}{}\begingroup
  \urlstyle{rm}\Url}\fi

\bibitem[{Bauso et~al.(2014)Bauso, Cannon, and Fleming}]{bauso2014robust}
Bauso, D., Cannon, M., and Fleming, J. (2014).
\newblock Robust consensus in social networks and coalitional games.
\newblock \emph{IFAC Proceedings Volumes}, 47(3), 1537--1542.

\bibitem[{Bauso and Notarstefano(2015)}]{Bauso2015}
Bauso, D. and Notarstefano, G. (2015).
\newblock {Distributed n-player approachability and consensus in coalitional
  games}.
\newblock \emph{IEEE Transactions on Automatic Control}, 60(11), 3107--3112.
\newblock \doi{10.1109/TAC.2015.2411873}.

\bibitem[{Blackwell(1954)}]{Blackwell1954}
Blackwell, D. (1954).
\newblock {An analog of the minimax theorem for vector payoffs}.
\newblock \emph{Pacific Journal of Mathematics}, 6(1), 1--8.
\newblock \doi{10.2140/pjm.1956.6.1}.

\bibitem[{Elsner et~al.(1992)Elsner, Koltracht, and Neumann}]{Elsner1992}
Elsner, L., Koltracht, I., and Neumann, M. (1992).
\newblock {Convergence of sequential and asynchronous nonlinear
  paracontractions}.
\newblock \emph{Numerische Mathematik}, 62(1), 305--319.
\newblock \doi{10.1007/BF01396232}.

\bibitem[{{Ernest Ryu} and Boyd(2016)}]{ErnestRyu2016}
{Ernest Ryu} and Boyd, S. (2016).
\newblock {a Primer on Monotone Operator Methods}.
\newblock \emph{Applied and computational mathematics}, 15(1), 3--43.

\bibitem[{Fele et~al.(2017)Fele, Maestre, and Camacho}]{fele2017coalitional}
Fele, F., Maestre, J.M., and Camacho, E.F. (2017).
\newblock Coalitional control: Cooperative game theory and control.
\newblock \emph{IEEE Control Systems Magazine}, 37(1), 53--69.

\bibitem[{Fullmer and Morse(2018)}]{Fullmer2018}
Fullmer, D. and Morse, A.S. (2018).
\newblock {A Distributed Algorithm for Computing a Common Fixed Point of a
  Finite Family of Paracontractions}.
\newblock \emph{IEEE Transactions on Automatic Control}, 63(9), 2833--2843.
\newblock \doi{10.1109/TAC.2018.2800644}.

\bibitem[{Han et~al.(2018)Han, Morstyn, and McCulloch}]{han2018incentivizing}
Han, L., Morstyn, T., and McCulloch, M. (2018).
\newblock Incentivizing prosumer coalitions with energy management using
  cooperative game theory.
\newblock \emph{IEEE Transactions on Power Systems}, 34(1), 303--313.

\bibitem[{Lehrer(2003)}]{lehrer2003allocation}
Lehrer, E. (2003).
\newblock Allocation processes in cooperative games.
\newblock \emph{International Journal of Game Theory}, 31(3), 341--351.

\bibitem[{Maschler et~al.(1971)Maschler, Peleg, and
  Shapley}]{maschler1971kernel}
Maschler, M., Peleg, B., and Shapley, L.S. (1971).
\newblock The kernel and bargaining set for convex games.
\newblock \emph{International Journal of Game Theory}, 1(1), 73--93.

\bibitem[{Nedich and Bauso(2013)}]{Nedic2013}
Nedich, A. and Bauso, D. (2013).
\newblock {Dynamic coalitional TU games: Distributed bargaining among players'
  neighbors}.
\newblock \emph{IEEE Transactions on Automatic Control}, 58(6), 1363--1376.
\newblock \doi{10.1109/TAC.2012.2236716}.

\bibitem[{Saad et~al.(2008)Saad, Han, Debbah, and Hj{\o}rungnes}]{Saad2009a}
Saad, W., Han, Z., Debbah, M., and Hj{\o}rungnes, A. (2008).
\newblock {A Distributed Merge and Split Algorithm for Fair Cooperation in
  Wireless Networks}.
\newblock \emph{IEEE Transactions on Wireless Communications}, 8(9),
  4580--4593.
\newblock \doi{10.1109/ICCW.2008.65}.

\bibitem[{Saad et~al.(2009)Saad, Han, Debbah, Hj{\o}rungnes, and
  Başar}]{Saad2009}
Saad, W., Han, Z., Debbah, M., Hj{\o}rungnes, A., and Başar, T. (2009).
\newblock {Coalitional game theory for communication networks}.
\newblock \emph{IEEE Signal Processing Magazine}, 26(5), 77--97.
\newblock \doi{10.1109/MSP.2009.000000}.

\bibitem[{Saad et~al.(2011)Saad, Han, and Poor}]{Saad2011}
Saad, W., Han, Z., and Poor, H.V. (2011).
\newblock {Coalitional Game Theory for Cooperative Micro-Grid Distribution
  Networks}.
\newblock In \emph{2011 IEEE International Conference on Communications
  Workshops (ICC)}, 1--5. IEEE.
\newblock \doi{10.1109/iccw.2011.5963577}.

\bibitem[{Scarf(1967)}]{scarf1967core}
Scarf, H.E. (1967).
\newblock The core of an n person game.

\bibitem[{Schmeidler(1969)}]{schmeidler1969nucleolus}
Schmeidler, D. (1969).
\newblock The nucleolus of a characteristic function game.
\newblock \emph{SIAM Journal on applied mathematics}, 17(6), 1163--1170.

\bibitem[{Shapley(1953)}]{shapley1953value}
Shapley, L.S. (1953).
\newblock A value for n-person games.
\newblock \emph{Contributions to the Theory of Games}, 2(28), 307--317.

\bibitem[{Smyrnakis et~al.(2019)Smyrnakis, Bauso, and
  Tembine}]{smyrnakis2019game}
Smyrnakis, M., Bauso, D., and Tembine, H. (2019).
\newblock Game-theoretic learning and allocations in robust dynamic coalitional
  games.
\newblock \emph{SIAM Journal on Control and Optimization}, 57(4), 2902--2923.

\bibitem[{Zolezzi and Rudnick(2002)}]{zolezzi2002transmission}
Zolezzi, J.M. and Rudnick, H. (2002).
\newblock Transmission cost allocation by cooperative games and coalition
  formation.
\newblock \emph{IEEE Transactions on power systems}, 17(4), 1008--1015.

\end{thebibliography}

\end{document}